# Application of Gauss' Law in Acoustics

Mladen Martinis [1] , Zoran Ozimec [2]

[1]*Rudjer Boskovic Institute, Zagreb, Croatia*
[2]*Braca Ozimec doo, Zagreb, Croatia*

April 3, 2011

## ABSTRACT

Practical application of Gauss' law in acoustics is not a very well known method. However, any inverse square law behavior can be formulated in the way similar to Gauss' law, which allows us to extend the same principle to sound waves propagation. We show in this paper how the acoustic power of sound source can be related to the sound intensity flow through a given surface by means of the Gauss' law. Several different sound-source shapes, important in practical applications, are analyzed by means of the Gauss' law. A suitable choice of the Gaussian surface allows us to obtain the simple and straightforward method for calculating the sound intensity distribution in space.

________________________________________________________________

[1]  martinis@irb.hr            [2]  post@zoran-ozimec.com

# 1. Introduction

Universal nature of Gauss' law [3] lies in the fact that it can be applied to almost any field having the inverse square law behavior in space. Obvious examples are gravity and electromagnetic fields, for which the Gauss' law gives information about the strength of the field sources enclosed by the Gaussian surface.

Consequently, applying Gauss' law to acoustics implies that sound wave flux through any closed surface will be related to the enclosed sound source. The sound source under consideration will be transducer [5], a device that converts electrical energy into acoustical energy. The minimal Gaussian surface will have the same shape and size as a radiating surface of the transducer, or the shape of the generated wavefront if those two are not the same. In this paper, we apply the Gauss' law directly to the effective (time averaged) flux of the sound intensity field.

We shall begin with the most basic example, the spherical sound source. By selecting the Gaussian surface also as the sphere but larger than sound source we can easily see how the inverse square law emerges.

Next example would be cylindrical sound source and Gaussian surface. Finally we shall show how it works for ellipsoidal source which is very close to real world example, the (curved) line array loudspeaker system [6],[7],[8]. In particular, it will be all horn line array system because horns acting as a waveguides allow multiple sound sources to integrate into single coherent one, making definition of sound source geometry and corresponding Gaussian surface practically straightforward.

The next Section is a brief introduction to how to use Gauss' law in acoustics.

## 2. Gauss' Law Details

The Gauss' law is a method widely used in electrical applications to calculate electric fields from symmetrically charged objects.

This law follows from the general divergence theorem applied to the continuity equation expressing the conservation of some physical quantity. In acoustic theory [1][2] the propagation of sound waves in air is described by two balance equations, that of mass and momentum, which, after some small-amplitude simplifications can be put in the form

$$\rho_0 \frac{\partial \mathbf{v}}{\partial t} + \nabla p = -\mathbf{f} \quad \text{(Momentum balance)} \tag{1}$$

$$\frac{\partial p}{\partial t} + \kappa \nabla \cdot \mathbf{v} = 0 \quad \text{(Mass balance)} \tag{2}$$

where $\rho_0$ is a constant denoting the average density of air, $p(\mathbf{x},t)$ is the air pressure field, $\mathbf{v}(\mathbf{x},t)$ is the velocity vector field of air particles, and $\mathbf{f}$ is an external driving force, a vibrating object, whose divergence $c_0^2 \nabla \cdot \mathbf{f} = s_p$ is considered as a source of sound pressure above the $p_0$-level. In addition, we must also take into account that this external perturbation is small and $\rho - \rho_0 \ll 1$ for sound waves.

Since air is a compressible fluid, the above equations of balance must be supplemented by an equation of state $p = p(\rho)$, relating the sound pressure to density. In this small-amplitude approximation, we can write $p(\rho) = p_0 + p'(\rho_0)(\rho - \rho_0)$, where $p'(\rho_0) = c_0^2$ is the square of the sound velocity in air at constant entropy. If the adiabatic equation of state $p = K\rho^\gamma$ of an ideal gas is valid, $c_0^2 = \gamma p_0 / \rho_0$. The bulk modulus of the air is denoted by $\kappa = \rho_0 c_0^2 = Z c_0$.

These two sets of balance equations can now be combined to a wave equation for the air pressure

$$\frac{\partial^2 p}{\partial t^2} - c_0^2 \nabla^2 p = s_p \tag{3}$$

whose solution is of the form

$$p(\mathbf{x},t) = p_0 + \int_{V_s'} \frac{s_p(\mathbf{x}',t')}{|\mathbf{x}-\mathbf{x}''|} d^3x' \qquad (4)$$

where $t' = t - |\mathbf{x}-\mathbf{x}'|/c_0$, and the integration is over the volume $V_s$ of the source. Similar solution can be found for the air-particle velocity field $\mathbf{v}(\mathbf{x},t)$ in the form

$$\mathbf{v}(\mathbf{x},t) = \int_{V_s'} \frac{\mathbf{s}_v(\mathbf{x}',t')}{|\mathbf{x}-\mathbf{x}'|} d^3x' \qquad (5)$$

where $\mathbf{s}_v = -\partial \mathbf{f}/\rho_0 \partial t$ is the particle velocity source.

The instantaneous acoustic intensity at a given point in space is defined as a product of the sound pressure and the particle velocity,

$$\mathbf{I}(\mathbf{x},t) = p(\mathbf{x},t)\mathbf{v}(\mathbf{x},t) \qquad (6)$$

The vector character of $\mathbf{I}$ shows in which direction the sound energy is flowing. The average acoustic intensity during the time interval $\Delta t$ is given by

$$\mathbf{I}(x) = \frac{1}{\Delta t}\int_0^{\Delta t} p(\mathbf{x},t')\mathbf{v}(\mathbf{x},t')dt' \qquad (7)$$

In acoustics the Gauss' law [3] becomes a powerful tool when its integral form is applied to the sound intensity vector in the form

$$\iiint_V \nabla \cdot \mathbf{I}\, dV = \oiint_S \mathbf{I}\cdot \mathbf{n}\, dS = P_{ac} \qquad (8)$$

where $P_{ac}$ is the acoustic power of the source enclosed by the Gaussian surface $S$. If the symmetry of the source is such that it allows to find the Gaussian surface on which

$$(\mathbf{I}\cdot \mathbf{n})|_S = (I_n)|_S = const. \qquad (9)$$

then we obtain a very simple formula, important in applications, for finding the sound intensity on the Gaussian surfaces

$$(I_n)|_S = \frac{P_{ac}}{S} = (I_n)|_{S_0} \frac{S_0}{S} \qquad (10)$$

where $S_0$ is the Gaussian surface very close to the sound source transducer.
In general, the Gaussian surface can be split in several smaller pieces on which the condition (9) is fulfilled. In that case we obtain the general formula

$$I_{n_1} S_1 + I_{n_2} S_2 + ... + I_{n_k} S_k = P_{ac} \qquad (11)$$

In the case of multiple coherent sound sources inside the Gaussian surface, the pressure waves should be added and resultant intensities calculated as $I = p^2/Z$ that is $I = (\sqrt{I_1} + \sqrt{I_2} + ... \sqrt{I_k})^2$. It is convenient to choose the closed Gaussian surface $S$ such that only one term on the left side of (11) dominates and others are considered as insignificant because of very low flux through them.

As far as frequency dependence is concerned it must be noted that for acoustically small sources (physical dimensions of source radiator are much smaller than the wavelength of radiated frequencies) the Gaussian surface shape and dimensions can differ significantly from those of the source radiator and in that case the shape of the generated sound wavefront must be considered.

From all that, it is clear that for any sound source of finite extension, the appropriate faraway Gaussian surface becomes spherical. Above mentioned practical considerations are out of the scope of this work and will not be analyzed here.

In the next section we shall analyze four cases of sound propagation that are important in practical applications. They correspond to a spherical source, a cylindrical source, a elliptic cylinder source and to a ellipsoidal source. All our results can be tested against the exact multipole solutions of (4), (5) and (6).

## 3. Applications

*Spherical source.* This is the simplest case, usually considered as a point source. In this case the Gaussian surface is a surface of the sphere with the radius $r$ so that $S_s = 4\pi r^2$ and using (10) we obtain a well known inverse square law

$$I_r = \frac{P_{ac}}{S_s} = I_{r_0}(\frac{r_0}{r})^2 \tag{12}$$

where $r_0$ is the radius of a spherical source and $r > r_0$.

*Circular-cylindrical source.* In this case, we use the cylindrical coordinates $(x = \rho\cos\varphi, \ y = \rho\sin\varphi, \ z = h)$, where $h$ is the height of the vertical height of the source, and $\rho$ is the radius of the cylinder. If $P_{ac}/h = const$ and $h$ is very large then (10) with $S_{cc} = 2\pi\rho h$ leads to

$$I_\rho = \frac{P_{ac}}{S_{cc}} = I_{\rho_0}\frac{\rho_0}{\rho} \tag{13}$$

Note, that $\rho = \sqrt{r^2 - h^2} \approx r$ for $r \gg h$ leads to $I_\rho \approx \frac{1}{r}$ behavior.

We also assumed that the sound energy flow through the top and the bottom surfaces of the cylinder could be neglected for a very thin and long source.

*Elliptic-cylindrical source.* Here we have a more complicated shape of the Gaussian surface given by the surface element

$$dS_{ec} = \sqrt{\rho^2 + (\frac{d\rho}{d\varphi})^2} d\varphi dz \tag{14}$$

where $\rho = \rho(\varphi) = a/\sqrt{1 - k^2\sin^2\varphi}$ and $k^2 = 1 - a^2/b^2$ with $a$ and $b$ being the semi-axes of the ellipse.

Integrating $dS_{ec}$ over $\varphi$ and $z$, we obtain $S_{ec} = 4bhE(k)$, where $E(k)$ denotes the complete elliptic integral of the second kind [4]

$$E(k) = a \int_0^{\frac{\pi}{2}} \frac{d\varphi}{\rho(\varphi)} \tag{15}$$

The equation (10) then reads

$$I_{ec} = \frac{P_{ac}}{S_{ec}} = I_{0ec} \frac{b_0 E(k_0)}{b E(k)} \tag{16}$$

*Ellipsoidal source.* In this case the Gaussian surface is an ellipsoid defined by three semi-axes $a$, $b$, and $c$ satisfying the equation

$$\frac{x^2}{a^2} + \frac{y^2}{b^2} + \frac{z^2}{c^2} = 1 \tag{17}$$

In spherical coordinates this surface is described by

$$r(\theta,\varphi) = abc(b^2 c^2 \sin^2\theta \cos^2\varphi + a^2 c^2 \sin^2\theta \sin^2\varphi + a^2 b^2 \cos^2\theta)^{-\frac{1}{2}} \tag{18}$$

The surface element of the ellipsoid is then $dS_{el} = r^2(\theta,\varphi)\sin\theta d\theta d\varphi$.

The integration over angular variables $(\theta,\varphi)$ gives us the whole surface $S_{el}$ of the ellipsoid that can be approximated by

$$S_{el} = 4\pi \left[ \frac{a^p b^p + a^p c^p + b^p c^p}{3} \right]^{\frac{1}{p}} \tag{19}$$

where $p = 1,6$. The normal component of the sound intensity follows the ellipsoidal orthogonal coordinate lines defined by confocal hyperboloids.

Sound intensity in the direction of the surface normal **n** is $I_n = \frac{P_{ac}}{S_{el}}$.

By an appropriate choice of the ellipsoidal semi-axis $a$, $b$ and $c$, we can describe our earlier examples as particular cases. We note here that $S_{el}$ in the form (19) allow us to obtain the sound intensity to behave with distance from the sound source as $I(r) \approx \frac{P_{ac}}{r^\alpha}$ where α can take all the values between zero and two.

## 4. Conclusion

We showed in this paper that use of Gauss' law in acoustics can be a powerful mathematical tool in practical applications.

We applied the Gauss' law to some of the basic sound-source shapes occurring in practice. They were the spherical, cylindrical and the ellipsoidal shapes, respectively. By choosing the Gaussian surfaces to be confocal to the basic shape, it can be easily seen how the inverse square law emerges at a very large distance from the source in each particular case. However, closer to source, we can see that inverse square law can attain any power of distance between 2 and 0. Two being for spherical source, one for cylindrical, zero for source producing plane waves, and for ellipsoidal anything between.

Modern large scale sound reinforcement systems are based on the line array loudspeaker concept pioneered by Heil and Urban [6] and later advanced by Staffeldt and Thompson [7],[8]. Their mathematical modeling of multiple sound sources were inspired basically on Huygens-Fresnel principle used in optics (integration of multiple directive point sources placed on the straight or curved line). Today's most advanced line array loudspeaker designs are based on waveguides-horns integrated into one virtually continuous area. Such design made possible to use more advanced mathematical tools like Gauss' law to describe sound radiation more precisely than it was possible with previous methods.